\documentclass[floatfix,twocolumn,aps,pre]{revtex4}
\usepackage{graphicx}
\usepackage{latexsym}
\usepackage{amsmath, amsthm, amssymb}
\usepackage{epsfig}
\usepackage{hyperref}
\usepackage{cleveref}

\begin{document}

\title{Dynamics of a polymer under multi-gradient fields}
\author{Sadhana Singh and Sanjay Kumar}
\affiliation{Department of Physics, Banaras Hindu University, Varanasi 221 005, India}

\begin{abstract}
Effects of multi-gradient fields on the transport of a polymer chain have been
investigated by using generalized Langevin dynamics simulations. We observe that the
natural frequency of tumbling follows $Wi^{0.66}$ scaling, where $Wi$ is the Weissenberg number. 
Analysis of angular tumbling time distribution reveals that the tail of distribution follows
exponential distribution and at high Weissenberg number, deviates from Poisson behaviour. 
Competition between velocity gradient which results shear flow in the system, 
and solvent quality gradient arising
due to the interaction among monomers revealed that there is another
scaling associated with the angular tumbling time distribution. 
Moreover, at low temperature, we observe unusual behaviour that at intermediate shear rates, decay rate 
$\nu$ decreases with $Wi$.  
\end{abstract}

\maketitle

Multi-factor gradients are essential components of many biological phenomena. They are 
responsible in  coordinating with one another to bring about cell-, time- and location- 
specific responses in living systems \cite{albert}. It is realized to be an important, 
evolutionarily conserved signalling mechanism for guiding the growth, migration, and 
differentiation of cells within the tissue \cite{Thomas}. They serve essential roles in inflammation, 
wound healing, cancer metastasis etc. Insight into the behavior of such systems is of fundamental 
importance in a wide spectrum of systems ranging from biological cells, where transport appears
in varying environments, to shear flow of biomolecules (including bacteria). The presence
of shear flow  arising due to velocity gradient affects the transport and dispersion 
of biomolecules at the macroscale \cite{Bird, Doi}. Theoretically, de Gennes long back showed that 
the behavior of coil-stretch transition in polymer is highly dependent on the type 
of flow \cite{degenes}. Later Smith and coworkers \cite{smith} monitored the 
motion of individual molecules (DNA) under the shear flow. 
They observed tumbling motion of a polymer chain under shear flow {\it i.e.}, 
a DNA undergoes a cyclic stretching and collapse dynamics, with a characteristic 
frequency which depends on the shear rate and its internal relaxation time. As a result, 
the dynamics of flexible polymers in shear flow  drew considerable interest in recent years 
\cite{schroeder, Buscalioni, Doyle, Winkler, victor, sanjib}. 

A single polymer chain in solution undergoes a transition from the coil (high 
temperature) state to the globule/folded (low temperature) state \cite{degenes1} 
as the temperature is lowered. It is also possible to study the coil-globule transition 
by changing the solvent quality {\it i.e.} by varying the interaction among monomers. 
A solvent is called a good solvent if polymer is found to be in the coil 
state, whereas it is referred as a poor solvent if polymer is in the globule state 
\cite{degenes1}. In addition to velocity gradient, one may think of a system having  gradient arising 
due to the varying  environment (solvent quality). In fact, chemotaxis is one of the process arising due 
to the change in solvent quality (chemical or concentration gradient), which leads to 
directional motion of cells, bacteria, biomolecules towards or away from a source \cite{pnas}. 
Notably, in experimental set-ups as well as in theories, one manipulates shear rate keeping 
the quality of solvent constant, thereby transport of biomolecules under shear flow is 
mainly explored by the velocity gradient \cite{smith, schroeder, Buscalioni, Doyle, Winkler, victor, sanjib}. 
The aim of this letter is to study the effect 
of net gradient field arising due to the competition between velocity and chemical potential 
on the dynamics of polymers, which still remains an unexplored territory. Since, the inclusion of 
gradient interactions to vary solvent quality in the model system under shear flow hampers 
the analytical treatment, therefore, we resort to computer simulations to shed light on the 
rich dynamical behaviour of such systems. 
\begin{figure}[t]
\includegraphics[width=0.5\textwidth]{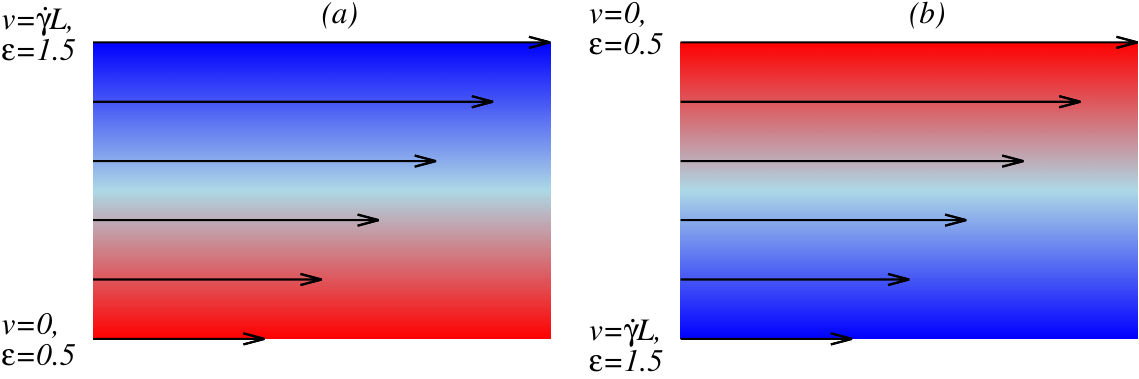}
\caption{Schematic diagram of flow profile; (a) with a positive gradient of interaction 
energy $\varepsilon$, both flow velocity, $v$ and interaction energy, $\varepsilon$ 
increase in the same direction, (b) with a negative gradient of $\varepsilon$ .}
\label{fig1}
\end{figure}

We have considered the composite system consisting of polymer of $N (= 50)$ beads (\cref{sec:supplementary}) 
and fluid 
(implicit solvent) confined between two walls such that one is stationary and other is moving with velocity 
$V_x$ resulting a velocity gradient in the flow, in the direction perpendicular to wall. 
The flow velocity experienced by each monomer is $\dot{\gamma}y_i$, 
where $\dot{\gamma}=\frac{dV_x}{dy}$, is the velocity gradient (shear rate) along the $y-$ direction. 
In order to exclude the effect of confinement, we have taken the width of channel 
greater than 4 time of radius of gyration of polymer.
To confine the system in a channel of width $L (=20)$, we have taken the particle-wall interaction
(at $y=0$ and $y=L$) in the form of soft repulsion of the Weeks-Chandler-Andersen potential \cite{deb}. 
Simulation is carried out in the the reduced units (\cref{sec:supplementary}).
Solvent quality gradient has been incorporated in the 
model system by linearly varying the interaction energy ($y\Delta\varepsilon$) associated with
non-bonded monomers along the $y-$ direction.  
Fig.\ref{fig1} shows the schematic of flow profile with positive and 
negative gradient of $\varepsilon$, where
value of $\varepsilon$ increases (Fig.\ref{fig1}(a)) and decreases (Fig.\ref{fig1}(b))
in direction of positive gradient of flow velocity, respectively. 
Here, we have taken $\Delta \varepsilon=0.05$ in the simulation. 
A change in color 
from red to blue (or {\it vice versa}) represent a gradient arising due to the change
in solvent quality (or temperature). The dynamics of the $i^{th}$ bead of polymer chain 
(\cref{sec:Supplementary}) in shear flow is described by 
the generalized Langevin equation (GLE), which explicitly 
takes into account the effect of coupling to a thermostat to maintain the constant temperature 
(\cite{mcphie, dobson1, dobson2} and \cref{sec:supplementary}).  

\begin{figure}
\includegraphics[width=0.5\textwidth]{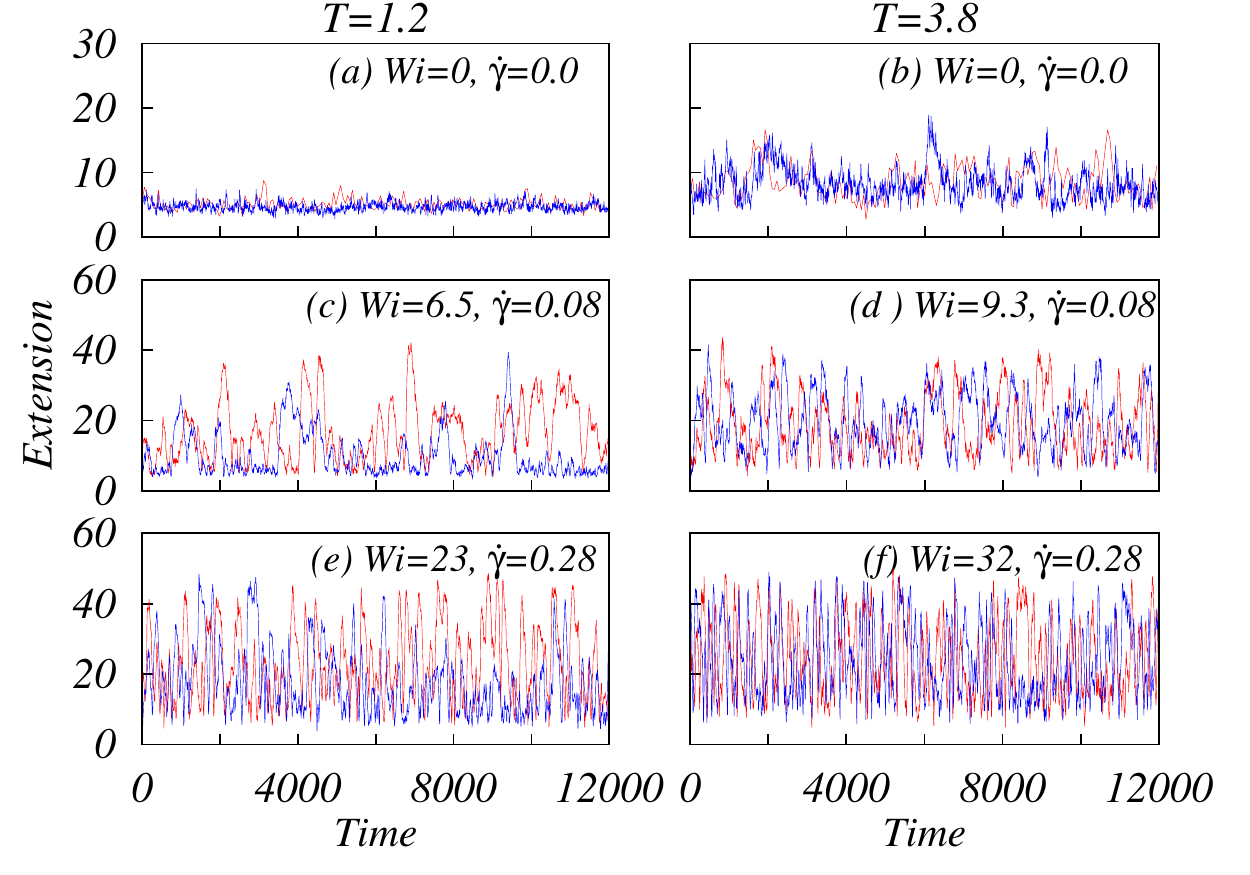}
\caption{Time history of extension of polymer chain in flow direction 
at different Wiessenberg number, $Wi$. (a), (c) and (e) at temperature T=1.2 (poor solvent); 
(b), (d) and (f) at T=3.8 (good solvent) for both cases $\varepsilon$=$Const$ (red color)
and $\varepsilon$=$y \Delta \varepsilon $ (blue color).} 
\label{fig2}
\end{figure}

In absence of flow, a decrease in temperature leads  to the coil-globule transition.
The ${\theta}-$temperature, $T_{\theta}$ at which this transition takes place, is estimated 
from the measurement of $\frac{<R_G^2>}{N}$ at different temperature, 
where $R_G$ is the radius of gyration. The $\theta$ temperature ($\varepsilon=1.0$) has been 
estimated to be $T_{\theta}=2.5 \pm 0.1$. At this temperature
$\frac{<R_G^2>}{N}$ is found to be independent of $N$ \cite{murat}. We have performed simulation 
in both regimes {\it i.e.} $T > T_{\theta}$ (good solvent) and $T<T_{\theta}$ (poor solvent).
The dimensionless flow strength is characterized by the Weissenberg number, 
$Wi = \dot{\gamma}\tau_0 $. Here, $\tau_0$ is the longest relaxation time of the polymer,
which has been determined by fitting the time decay of the autocorrelation 
function of the end-to-end distance of polymer chain in the absence of a solvent flow.
In Fig.\ref{fig2}, we have compared the time series of extension of a single chain
at different $Wi$ for the uniform solvent quality ($\varepsilon= constant$) and the varying
solvent quality ($\varepsilon$=$y \Delta \varepsilon$). Fig.\ref{fig2}(a) and (b) show
the equilibrium extension ($\dot\gamma =0$) of polymer chain in poor ($T= 1.2 < T_{\theta}$) and good
($T=3.8 > T_{\theta}$) solvents, respectively. It is evident from the Fig.\ref{fig2}(c) that
at low shear rate, polymer chain for varying interaction remains in globule state for a longer 
time compared to that of the constant interaction ($T < T_{\theta}$). However, at high temperature 
($ T >T_{\theta}$) there is rapid fluctuations in the extension  (Fig.\ref{fig2}(d)) for both 
types of solvent. The maximum extension of polymer depends on the shear rate irrespective 
of solvent quality. At high shear rate,  (Fig.\ref{fig2}(e) and (f)) show significant
increase in the tumbling events.

\begin{figure}[]
\includegraphics[width=0.5\textwidth]{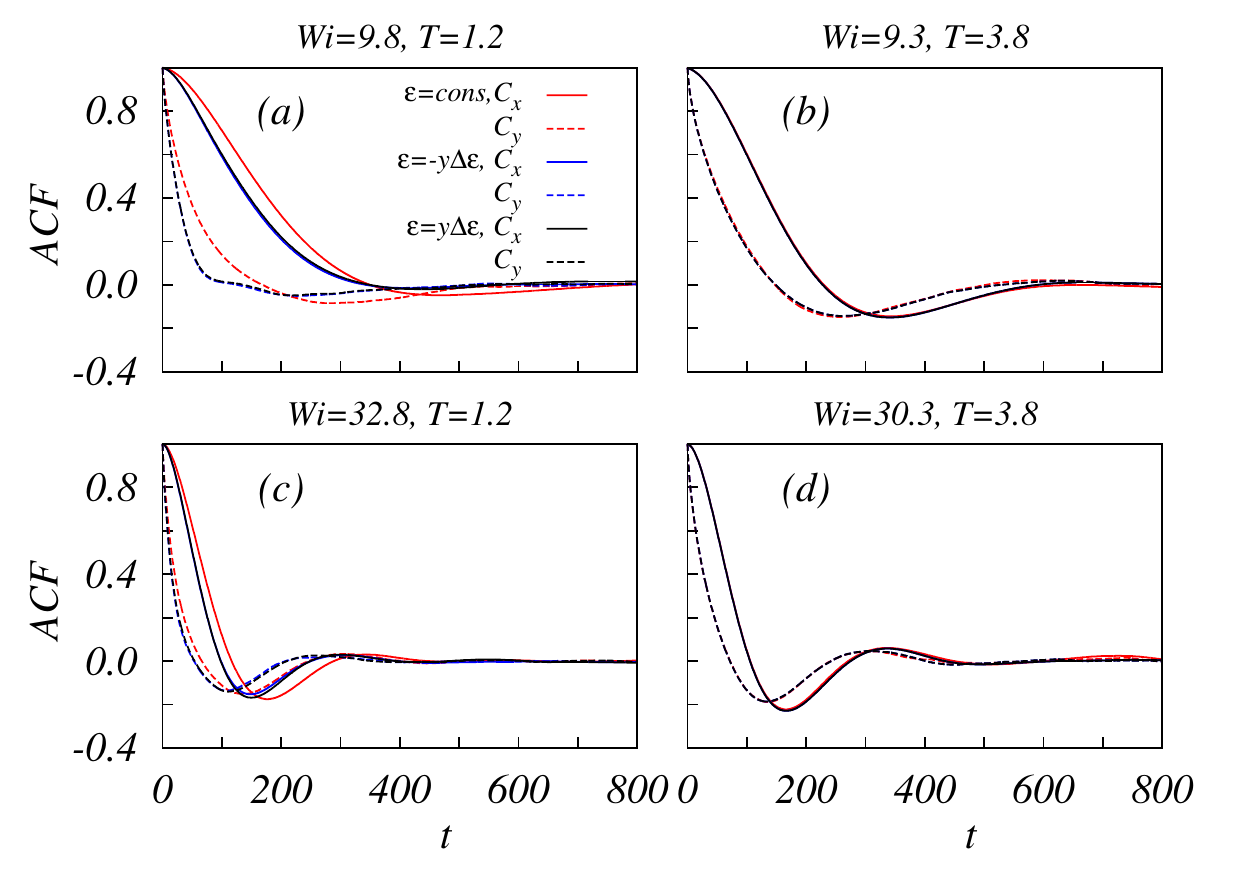}
\caption{Normalised autocorrelation function (ACF) 
of the components of end to end vector $\vec{R}$ in 
flow and gradient direction $C_x$, $C_y$ for both cases 
$\varepsilon$=$cons$ and $\varepsilon$=$y \Delta \varepsilon $. 
(a) and (c) at different $Wi$ at temperature $T$=$1.2$, 
for poor solvent condition; 
(b) and (d) at $T$=$3.8$ for good solvent condition.}
\label{fig3}
\end{figure}

To study the tumbling dynamics, we calculated 
the autocorrelation function (ACF) $C_x$, $C_y$ \cite{Schroeder1,Buscalioni,Florencio} of components 
$x$ and $y$  of end to end vector, $\vec{R_e}$ in flow and gradient direction respectively, 
which are defined as $C_{\alpha}(t) = \langle{ \delta{R_{\alpha}(t)} \delta{R_{\alpha}(0)}}\rangle /
\langle{ \delta{R_{\alpha}(0)} \delta{R_{\alpha}(0)}}\rangle$, where $\alpha = x, y$, 
$\delta{R_{\alpha}(t)}=R_{\alpha}(t)-\langle{R_{\alpha}}\rangle$ and $\langle{.}\rangle$ 
denotes the time average. 
Fig.\ref{fig3} shows the ACFs $C_x$, $C_y$ for both cases $\varepsilon$=$constant$ and 
$\varepsilon$=$y \Delta \varepsilon $ at temperatures $T=1.2$ and $T=3.8$. 
One remarkable feature which can be noticed from these plots is that the tumbling dynamics 
of polymer chain remains insensitive to the positive and negative interaction gradients. 
Furthermore, the effect of interaction gradient is visible only 
in the case of poor solvent at low $Wi$. 
At higher $Wi$, this effect is 
vanishing and both the ACFs behave similar to the case of
$\varepsilon=constant$ (Fig.\ref{fig3}(c)).
In a good solvent ($T>T_{\theta}$), there is no effect of interaction gradient on ACFs 
even at lower value of $Wi$ (Fig.\ref{fig3}(b) and (d)).   

\begin{figure}[]
\includegraphics[width=0.5\textwidth]{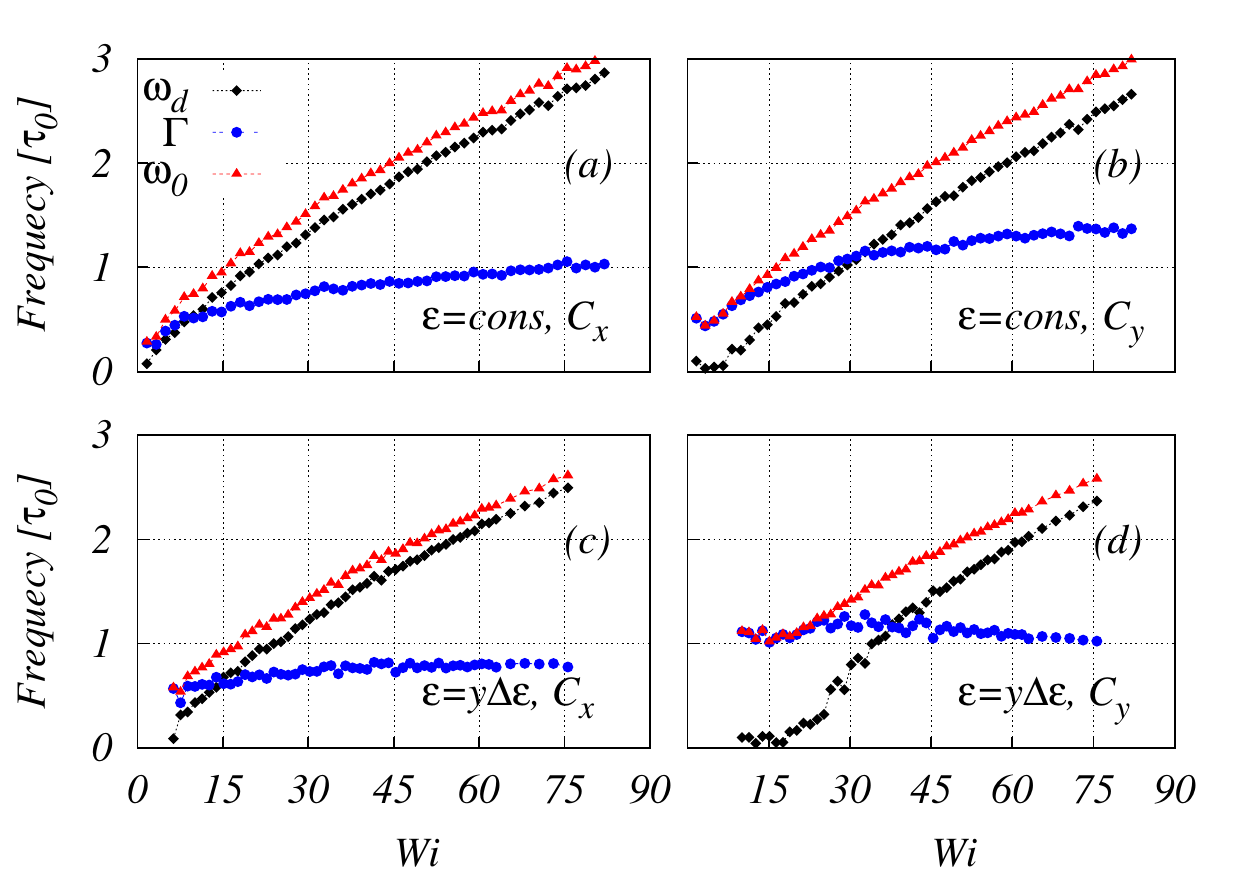}
\caption{Values of $w_d$, $\Gamma$ and $w_0$ obtained from fits 
to ACF ($c_x$ and $c_y$) for T=1.2 (poor solvent). }
\label{fig4}
\end{figure}

The dynamics of the chain under shear flow is characterised by well defined tumbling events. 
It is dissipative in nature and arising due to the external forcing similar to the one seen in
randomly excited damped oscillator. This analogy may be used to calculate the characteristic 
time involved in tumbling. The generic form of a damped harmonic oscillator  
$F(t)=A^2 cos(w_dt+ \psi)exp(-\Gamma t)$ was used to fit the ACFs, $C_x$ and $C_y$ \cite{Florencio}.
The damping rate $\Gamma$, the natural frequency $w_0$ ($w_o^2$=$w_d^2+\Gamma ^2$)
and the phase constant $\psi$ are the three time parameters. 
The phase lag $\Delta \psi $=$\psi _y-\psi _x $ is always 
positive and gives a new characteristic time 
$\tau _{lag} = \Delta \psi/w_d$, which indicates how fast the chain extension $X$ in 
shear direction responses to a drag force arising due to the fluctuation in the extension
$Y$ in the gradient direction. 
It is evident from Fig.\ref{fig3}(a) that there is an increase in $\tau _{lag}$ at low $Wi$ 
due to the interaction gradient at low temperature. Values of $w_d$, $\Gamma$ and $w_0$ obtained 
from the fits of  ACFs ($C_x$ and $C_y$) for $T=1.2$ (poor solvent) are shown in Fig.\ref{fig4} \cite{good}. 
There are significant differences in the values of parameters of both ACFs.
Furthermore, the dynamic of chain in flow direction is underdampped ($w_0 \simeq w_d$). However, for low value of 
$Wi(< 30)$, one observes $w_0 \simeq \Gamma$
in gradient direction. This difference becomes more prominent in presence of 
interaction gradient {\it i.e.} $w_0 \simeq \Gamma$ and $w_d \simeq 0$, 
while dynamics in flow direction remains unaffected. 
In all cases, the motion of the chain becomes underdamped as $Wi$ increases
($w_0 \simeq w_d$ and $\Gamma << w_0$). The natural frequency $w_0$ is related to
tumbling time for tumbling process, succession of coil-stretch cycle, $\tau _{tumb}=\pi /w_0$
\cite{Florencio}. In all cases, irrespective of directions, good or poor solvent quality,
absence or presence of interaction gradient,
the natural frequency is found to be scaled as $Wi^{2/3}$ for high $Wi$, 
a robust feature of tumbling dynamics found in previous theoretical and experimental studies
\cite{schroeder,victor,Florencio}.

\begin{figure}[t]
\includegraphics[width=0.4\textwidth]{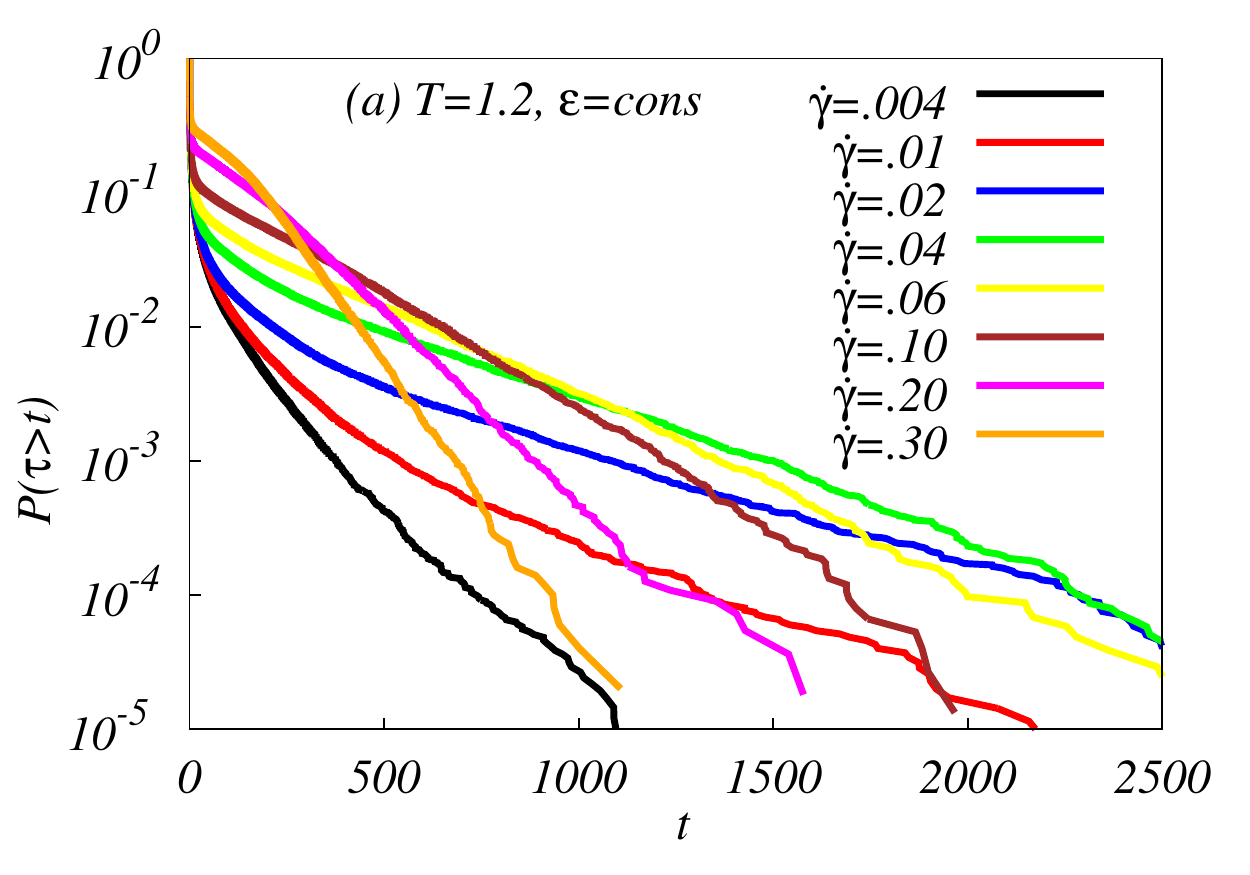}\\
\includegraphics[width=0.4\textwidth]{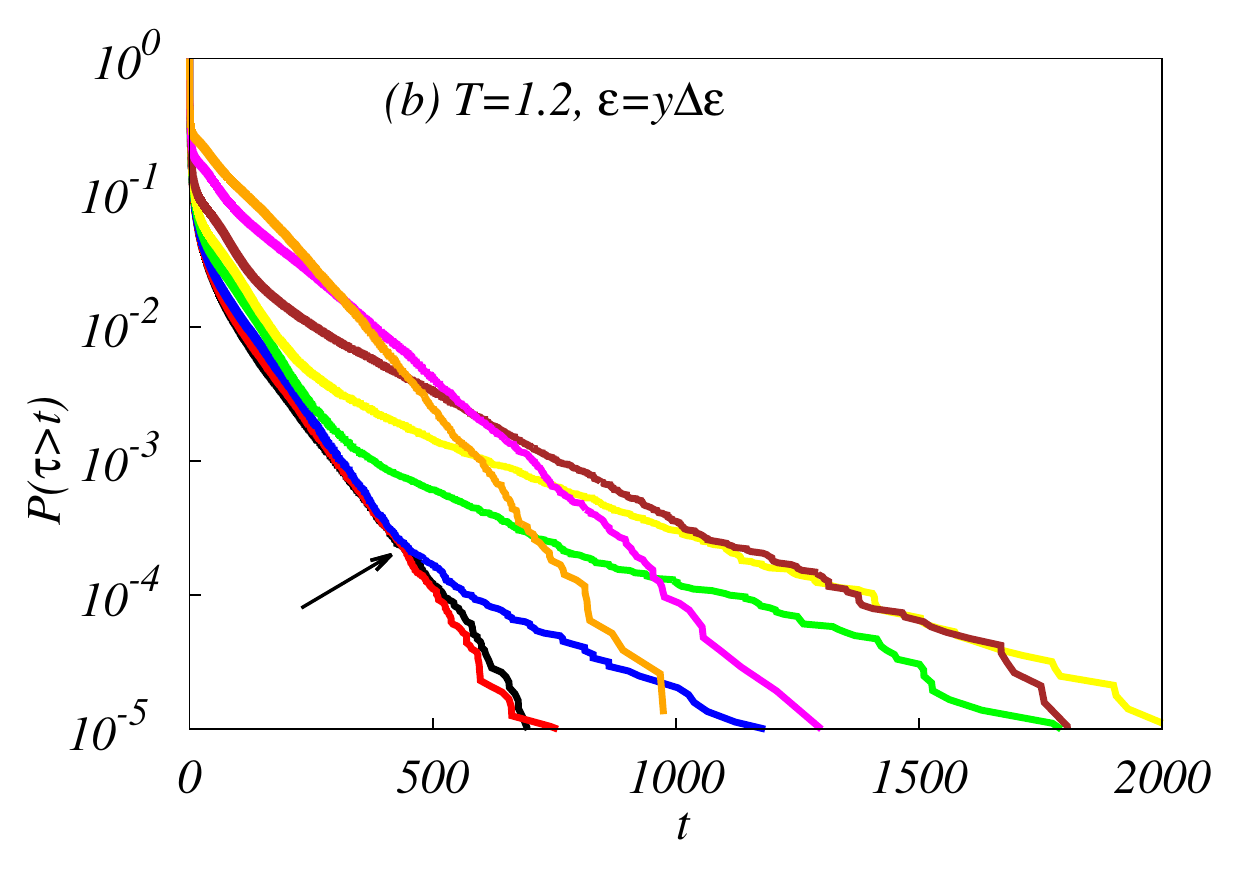}
\caption{Cumulative distribution of characteristic tumbling time $\tau$ 
, at different $Wi$ for both cases (a) $\varepsilon$=$cons$ and 
(b) $\varepsilon$=$y \Delta \varepsilon $, at temperature $T$=$1.2$, 
for poor solvent condition. Arrow demarcates the emergence of new  scaling, which is absent in (a).} 
\label{fig5}
\end{figure}

Tumbling of chain in shear flow is a stochastic process \cite{smith,Teixeria,Schroeder1}. 
The tumbling time, i.e. time interval between subsequent flips of the polymer, is a
random variable with relatively large fluctuations.  Now onwards, we focus our study on 
the distribution of angular tumbling time $P(\tau)$, where $\tau$ is the time interval
between two subsequent zero crossing of end-to-end distance, $R_x=x_n-x_1 $ in the flow direction.
For sufficiently large time intervals, it follows exponential distribution, 
$P(\tau) \approx exp(-\nu \tau)$ (Fig.\ref{fig5}), where {$\nu$} is the decay rate, its inverse gives
the information of $\tau_{tumb}$. This is in agreement with previous studies 
\cite{victor,celani, puliafito, sanjib, Florencio}. Fig.\ref{fig5}, shows the
distribution of angular tumbling time for different shear rates at low temperature. 
One can notice a change in slope at shear rate $\approx 0.02$ ($\varepsilon$ = constant) and $0.06$ 
($\varepsilon = y \Delta \varepsilon$) in Fig.\ref{fig5}(a) and (b), respectively. We 
identified these values as the critical shear rate, 
where polymer undergoes a shear induced coil-stretch transition 
\cite{katz}. 
The most interesting feature of the Fig.\ref{fig5}(b) 
is the presence of two time scales 
at intermediate shear rates for $\varepsilon = y \Delta \varepsilon$, which is absent for $\varepsilon$ = constant.

\begin{figure}[t]
\includegraphics[width=0.5\textwidth]{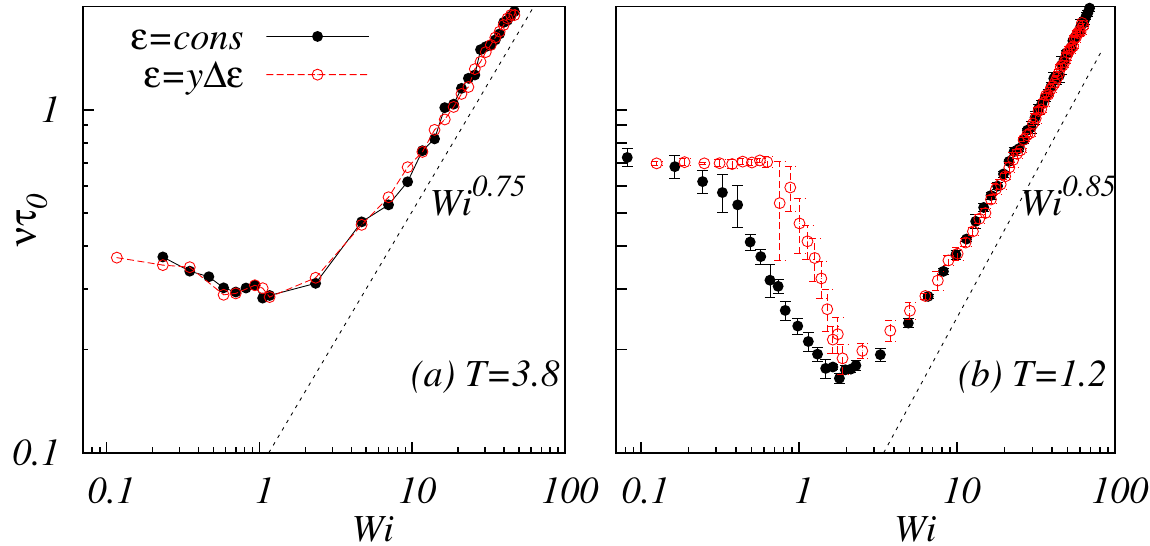}
\caption{The limiting decay rate, $\nu$ of tumbling time distribution scaled with 
relaxation time $\tau_0$. } 
\label{fig6}
\end{figure}

Fig.\ref{fig6}(a) and (b) show limiting decay rate $\nu$ of tumbling time distribution 
scaled with relaxation time $\tau_0$ with $Wi$ in poor and good solvent respectively. 
$\tau_ {tumb}$ is nearly independent of $Wi$ at low values of $Wi$ and proportional to $\tau_0$. 
At higher $Wi$, $\nu \tau _{0} \approx Wi^{0.75 \pm 0.05}$ and 
$\nu \tau _{0} \approx Wi^{0.85 \pm 0.05}$ at high and low temperature, respectively.
These exponents are steeper, show non-poissonian behaviour of tumbling process, in agreement with 
value in \cite{Florencio}. 
Surprisingly, at intermediate shear rates (below the critical shear rate), the decay rate decreases 
with  $Wi$. Moreover for $\varepsilon = y \Delta \varepsilon$, the decrease in decay rate is more 
steeper.  

In this letter, we have studied for the first time effects of multi-factor gradients on the 
tumbling dynamics of polymer chain in the free draining limit. We have incorporated 
the gradient of non-bonded interaction energy parameter, $\varepsilon$ associated with  LJ-potential 
as well as shear flow (velocity gradient) in the model system. 
In presence of the shear flow, two interesting cases namely positive 
and negative gradient of $\varepsilon$ with respect to velocity gradient were studied. 
Surprisingly, there are no notable differences in the tumbling dynamics of chain for both cases. 
Another interesting observation is
that the tumbling dynamics remains invariant for both cases $\varepsilon$ = constant and 
$\varepsilon = y \Delta \varepsilon$ for poor and good solvent in high shear regime. 
This is because the energy scale associated with tumbling due to rotation at higher shear rate is much higher 
than the energy needed to stabilize the polymer conformation. 
At lower shear rate in poor solvent, where both energy scales are comparable, 
one observes significant difference in ACFs ($C_x$, $C_y$), whereas in the good solvent
such difference is found to be absent. It is found that the tumbling 
frequency scales as $Wi^{2/3}$ (Fig.4)  consistent with previous studies. 
Similar scaling is reported if one considers hydrodynamics interaction \cite{Florencio, schroeder}.
Here, we show that it is 
the robust feature of the tumbling process irrespective of solvent quality, directions and 
gradient of chemical potential.  

The long tail behavior of tumbling time distribution is found 
to be exponential with non-poissionian exponent (Fig.6). The exponent for poor solvent is found 
to be steeper than that of the good solvent. 
At low shear rates, tumbling time, $\tau_ {tumb}$ is nearly independent of shear rate, 
At very low shear rates, the polymer
conformation will be in the collapsed state and its shape is nearly like a sphere 
of radius $R_G$. In this regime, sphere will roll and the work done due to shear force is
negligible compare to the interaction energy of polymer  beads. As a result, $\tau_ {tumb}$ 
is independent of shear rates. 
Above the critical shear rate, 
the work done by shear force is dominant and governs the tumbling dynamics of polymer. 
Increase in the rotational component of shear flow causes frequent tumbling of 
polymer accounting the decrease in $\tau_ {tumb}$ with increase in shear rate. 
The most striking feature of present study is the change in the trend of limiting decay 
rate below the critical shear rate at low temperature.  There is decrease in $\nu$ with 
increase in shear rates. At intermediate shear rate below the critical value, the work done 
due to shear force now is comparable to the interaction energy, and thus polymer deforms from the 
sphere like shape to an ellipsoid pointing along the flow direction, which inhibit tumbling. 
At this point, our work calls for further investigation by including hydrodynamics interaction 
and effects of chain size on the tumbling dynamics \cite{katz}.

It may be difficult to implement varying solvent quality in terms of the interaction gradient 
{\it in vitro}  similar to the one seen {\it in vivo}. However, if one maintains the two 
surfaces of the channel at two different temperatures, it is possible to achieve the steady state in 
temperature, which may be thought similar to the interaction gradient. Another way to realize
such a gradient {\it in vitro} could be due to the differences in the affinity of the salt 
with the top and bottom surfaces of the channel, which will give rise a salt gradient. One may 
consider the case of an ionic salt, where electric field gives rise to such gradient. 
Rheological studies involving isotropic molecules (spherical in shape) and anisotropic molecules 
(e.g. liquid crystals) may confirm our findings \cite{maren1, Cates, Foglino}.
Therefore, 
the present studies open several new issues, which warrant further experimental investigations 
to explore such hitherto unknown scaling related to the tumbling dynamics of polymer 
under multi-gradient fields, which may have potential applications in 
understanding the dynamics of active particles.  

We thank Garima Mishra and Apratim Chaterjee for many helpful discussions on the subject. 
We are grateful to P.J. Daivis and Matthew Dobson for their discussions on generalized Langevin equation 
for this study. The financial
assistance from SERB and INSPIRE program of DST, New Delhi, India are gratefully acknowledged.

\appendix
\label{appendix}

\section{Supplementary material}
\label{sec:supplementary}
In the present study, polymer chain is modelled by conventional bead-spring model \cite{kremer} 
consisting of N beads where the all non-bonded beads interaction causing excluded volume 
effect is given by 12-6 Lennard-Jones potential: 
\begin{equation}
\label{v_lj}
V_{lj}\left({r} \right) = 4\varepsilon \left(\left(\frac{\sigma}{r}\right)^{12}-\
            \left(\frac{\sigma}{r}\right)^6 \right),
\end{equation}
where, $r$ is the distance between two monomers, $\varepsilon$ and $\sigma$ characterize 
the strength of the interaction (potential) and the diameter of monomers, respectively. 
Throughout the simulation, we have worked in the reduced 
units, where $\varepsilon=1.0$, $\sigma=1.0$ and  $mass=1.0$ are 
the units of energy, distance and mass of beads, respectively. Units of other quantities are derived in 
terms of these parameters. Time is measured in units of 
${\sigma \left({\frac{m}{\varepsilon}}\right)^{\frac{1}{2}}}$. 

All bonded beads experience the finite extensible non linear (FENE) potential and
repulsive part of the 12-6 Lennard-Jones(LJ) potential, are given as
\begin{equation}
\label{pot_fene}
  V_{fene} = -\frac{kR_0^2}{2}\ln \left( 1- \left( \frac{r}{R_0} \right)^2 \right),
\end{equation}
\begin{equation}
\label{v_bond}
  V_{lj}^{b}\left({r} \right) = 4\varepsilon _{b}\left(\left(\frac{\sigma _{b}}{r}\right)^{12}-\
            \left(\frac{\sigma _{b}}{r}\right)^6 \right).
\end{equation}
where, $k$ is spring constant and $R_0$ is the maximum extension of bond.
We set $\varepsilon _{b}=1.0 \varepsilon$, 
$\sigma _{b}= 1.0\sigma$ and the values of other parameters are chosen 
from Ref. \cite{kremer} in which $R_0={1.5 \sigma}$, $k =30 \frac{\varepsilon}{\sigma^2}$.    

The cut-off distance of LJ-potential has been implemented by using the following definition of shifted 
force potential $V_{lj}^{sf}\left({r} \right)$ \cite{allen} to remove the
discontinuity in the potential as well as in the force: 
\begin{equation}
\label{v_lj_sf}
V^{sf}\left({r} \right)= 
\begin{cases}
V\left({r} \right) - V\left({r_{c}} \right) - \
\left({\frac{dV\left({r}\right)}{dr}}\right)_{r_{c}} \left( {r-r_{c}}\right), r \leq r_{c}\\
0,~~~~~~~~~~~~~~~~~~~~~~~~~~~~~~~~~~~~~~~~~~ r > r_{c} 
\end{cases}
\end{equation}
where $r_c$ is the cut-off distance for intermolecular interaction. 
We set $r_c=2.5\sigma$ and $2^{\frac{1}{6}}\sigma_b$ in Eqs.(\ref{v_lj}) 
and (\ref{v_bond}), respectively. 

The particle-wall interaction (at $y=0$ and $y=L$) of composite system (polymer and fluid) confined 
between two walls of width $L$ is taken in the form of soft repulsion of the Weeks-Chandler-Andersen 
potential \cite{deb}. For the wall at $y=0$,
\begin{align}
   V_w = 4\varepsilon_w \left( \left( \frac {\sigma_w}{y} \right)^{12} - \
          \left( \frac {\sigma_w}{y} \right)^6 + \frac {1}{4} \right),  \
          0 \leq y \leq 2^{\frac{1}{6}}\sigma_w .
\end{align}
For the wall at $y=L$,
\begin{align}
V_w &= 4\varepsilon_w \left( \left( \frac {\sigma_w}{L-y} \right)^{12} - \
          \left( \frac {\sigma_w}{L-y} \right)^6 + \frac {1}{4} \right), ~~\text{for} \nonumber \\
&~~~~~~~~~~~~~~~~~~~~~~~~~~~~~~~~L-\sigma_w \leq y \leq 2^{\frac{1}{6}}\sigma_w
\end{align}

Here, $y$ is the y-coordinate of the bead along the flow gradient. In present 
simulations, we choose $\sigma_w= \frac{\sigma}{2}$ and the parameter $\varepsilon_w$ controls 
the strength of soft repulsion. 

The dynamics of the $i^{th}$ bead of polymer chain in shear flow is described by the following Langevin equation:
\begin{equation}
 \vec{\ddot{r}}_i(t)=\vec{F}_i^c(t) - \Gamma \left[\vec{\dot{r}}_i(t) \
 - \dot{\gamma} y_i\hat i \right ] + \dot{\gamma}\dot{y_i}\hat i + \vec{R}_i(t)
\end{equation}
This is the generalized Langevin equation (GLE) for a particles in a steady shear flow, which explicitly 
takes into account the effect of coupling to a thermostat to maintain the steady state 
\cite{mcphie, dobson1, dobson2}. In this equation, the Langevin thermostat is applied to the relative velocity 
coordinate and then transformed back to the laboratory frame. 
The first term is the conservative force, $\vec{\triangledown}V$, $V=V_{FENE}+V_{LJ}$.
The second and third term  represent drag force relative to the flow velocity and 
force in the x-direction due to the particle's movement across the streamlines, respectively. 
The last term is the random force acting on $i^{th}$ bead due to solvent motion  is \cite{kubo, mcphie, dobson1, dobson2}:
\begin{equation}
\langle{{\vec{R}_i(t) \vec{R}_j(t')}}\rangle = {6K_B T \Gamma \delta_{ij} \delta(t- t')}.
\end{equation}

Eq.(A7) is solved by using a fifth-order predictor corrector method \cite{allen}, 
with a time step $\Delta t=0.006$.  
The simulation has been carried out over a broad range of friction coefficient 0.5 to 8.0.
In order to vary $Wi$, one has to change either friction coefficient or shear rate. 
For a given shear rate, 
we vary friction coefficient to achieve desire range of $Wi$. We have checked the 
stability of simulation for long run.
We have calculated the temperature from thermal momentum numerically and 
found that  the temperature of the system remains same as of the bath temperature throughout the simulations. 


\begin{thebibliography}{}
\bibitem{good} In case of good solvent, the values of $w_d$ and $\Gamma$ for both ACFs are 
found to be equal within the error bar.
\end{thebibliography}

\end{document}